\documentclass{ws-ijmpe}

\begin{document}

\markboth{Ernesto Kemp, Bruno Miguez and Walter Fulgione}{Estimating the distances of stellar collapses in the galaxy analyzing the energy spectrum of neutrino bursts}

\catchline{}{}{}{}{}
%%%%%%%%%%%%%%%%%%%%%%%%%%%%%%%%%%%%%%%%%%%%%%%%%%%%%%%%%%%%%%%%%%%%

\title{Estimations of the distances of stellar collapses in the galaxy by analyzing the energy spectrum of neutrino bursts}

\author{Ernesto Kemp, Bruno Miguez}

\address{Instituto de F\'{i}sica ``Gleb Wataghin'', Universidade Estadual de Campinas\\
Campinas, SP 13083-859, Brasil\\
kemp@ifi.unicamp.br}

\author{Walter Fulgione}

\address{ Istituto di Fisica dello Spazio Interplanetario IFSI-INAF and\\
Istituto di Fisica Nucleare – INFN, Italy\\}

\maketitle

\begin{history}
\received{(received date)}
\revised{(revised date)}
%\accepted{(Day Month Year)}
%\comby{(xxxxxxxxxx)}
\end{history}

\begin{abstract}
The neutrino telescopes of the present generation, depending on their specific features, can reconstruct the neutrino spectra from a galactic burst.
Since the optical counterpart could be not available, it is desirable to have at hand alternative methods to estimate the distance 
of the supernova explosion using only the neutrino data. In this work we present preliminary results on the method we are 
proposing to estimate the distance from a galactic supernova based only on the spectral shape of the neutrino burst and assumptions on the gravitational 
binding energy released an a typical supernova explosion due to stellar collapses.
  
% There is a consensus in both theoretical basis as well as from observations coming from the SN1987A that supernova 
%explosions are preceded by a huge flux of neutrinos from six species. In principle, large neutrino telescopes are able to 
%identify the burst signal from the background fluctuations and them to carry on further analysis over data and, where applicable, 
%build the energy spectrum from $\bar{\nu_e}$, providing astrophysical information from the innermost regions of the collapsed object. 
%In this work we show some preliminary results about the method we are proposing to extract the distance from the collapse, using only the 
%information embedded in the energy spectrum of expected $\bar{\nu_e}$, events on neutrino telescopes, such as the LVD at the LNGS - Italy, 
%and assumptions on the total binding energy of the stellar core released in the collapse process. In this preliminary phase we focus the discussion 
%on the errors of the method related to the theoretical uncertainties concerning the energy released in the formation of the compact object resulting 
%from the collapse.
\end{abstract}

\section{Introduction}
There is a consensus both in theoretical\cite{0} and experimental frameworks, the latter coming from observations of the 
SN1987A\cite{SN1987A-1,SN1987A-2,SN1987A-3,SN1987A-4}, that a supernova explosion is preceeded by a huge neutrino flux of
all flavours and their antiparticles. 
In principle, large neutrino telescopes are able to disentangle the neutrino burst signal from the background fluctuations, 
and to sample data with statistical significance to build the neutrino energy spectrum.
The method proposed in this work is dedicated to estimate the distance from the collapsing star, i.e. the supernova explosion, by using 
only the information of the spectral shape from expected $\bar{\nu_e}$ events on neutrino telescopes, as well as 
assumptions on the total binding energy of the stellar core released in the collapsing process. In this preliminary phase our discussion is 
focused on the errors of the method related to the theoretical uncertainties concerning the energy released in the formation of the compact 
object from the collapsing star.

\section{Motivation}
The distance from a galactic SN explosion determined by other methods can present some difficulties, for instance, 
the optical counterpart from galactic neutrino bursts can be fainted, introducing large errors in traditional photometry,
or even obscured by the disk matter, making measurements in the visible spectrum completely unfeasible. Triangulation with 
other neutrino telescopes could result in large angular errors. Parameter scanning in data analysis is a possibility. 
In this case, the result of this method can be used as an independent input, or even as a prior in other statiscal methods such
as a bayesian approach.
%. The detection of galactic neutrino bursts can become harder due the obscuring of the optical signal by the disk matter and triangulation with some neutrino detector can have too large errors.\par
%Detection of galactic neutrino bursts can have their optical counterpart obscured by the disk matter.
%Triangulation with other neutrino telescopes could have large errors.
%Parameter scanning in data analysis is a possibility. In this case, the result of this method can be used as an independent input, or as a prior.

\section{The Method}
The number of expected events in a neutrino telescope, $N_\nu$, is given by the expression:
\begin{equation}
 \label{numeventos}
N_\nu= \frac{N_T}{4 \pi D^2} \int_{E_{th}}^{\infty} \frac{dN_\nu}{dE} \sigma(E) \epsilon(E) dE
\end{equation}
Where $N_T$ is the number of target particles, $\frac{dN_\nu}{dE}$ is the SN neutrino spectrum, $\sigma(E)$ is the cross-section of the 
neutrino interaction to be detected and $\epsilon(E)$ is the detection efficiency. D is the SN distance we are trying to estimate from data analysis.
Supernova neutrino spectra are generally calculated by numerical methods and were obtained by different groups. We adopt the most standard 
approach, a Fermi-Dirac spectrum with a temperature T and a non-thermal parameter $\eta$ (pseudo-degeneracy) \cite{1}:
\begin{equation}
 \label{espectro}
\frac{dN_\nu}{dE}= \frac{AE^2}{1+e^{x-\eta}}
\end{equation}
Where $x=E/T$ and $A = E_{tot}/T^4f^3$ the normalization imposed by the total energy  $E_{tot}$  emitted as neutrinos, $f^3$ is the Fermi 
integral of 3rd order.\par

An important issue of the method is that $E_{tot}$ can be related to the binding energy $E_b$ released in stellar collapse, 
since 99\% of $E_b$ is carried away by neutrinos.
The binding energy $E_b$ is given by\cite{2}: $ E_b \approx 1.60 \times 10^{54}erg  \frac{M_{core}}{M_{\odot}}  \frac{10km}{R_{NS}}$.
Considering that $M_{core}$ is tightly bound to the Chandrasekhar mass (1.2-1.4)$M_\odot$ and $R_{NS}$ is the neutron star radius (12 - 15 km), 
thus $E_b$ is within a rather narrow range and consequently $E_{tot}$ is constrained between $2.0-3.0 \times 10^{53} erg$. 
From Eq.\ref{numeventos} one can see that an observation of a given
number of counts by an experiment result as information only the ratio $A/D^2$, i.e the normalization 
of the total neutrino flux diluted by the square of the collapse distance. We can go further if only from the $\bar{\nu_e}$ spectral shape the 
parameters (T,$\eta$) could be determined. Since $E_{tot}$ is constrained as described above, by assuming a 
standard value for $E_{tot}$ within a conservative range, it is possible to obtain the propper normalization constant 
for Eq.\ref{espectro} rather accurately.
Then, we can solve Eq.\ref{numeventos} for the distance $D$, as it is the last free parameter.\par
The method can be summarized as follows: \textbf{1.} A neutrino telescope measures $\bar{\nu_e}$ events and builds their energy spectrum.
\textbf{2.} The Kolmogorov-Smirnov test is applied over data to determine spectral parameters 
(T, $\eta$). KS-test was specially chosen since is a test for shapes and it does not depend on relative normalizations of the distributions,
thus at this step, $E_{tot}$ and D are not relevant. \textbf{3.} Counts from step 1. and parameters from data analysis (step 2.) are then
used as inputs on the Eq.\ref{numeventos}, which is solved for D, by assuming as a prior the conservative, but
nonetheless restrictive assumption $E_{tot} = 3,0 \times 10^{53}$ erg  
as discussed before.  We expect that the narrow range in $E_{tot}$ can have minimum impact over the 
distance estimation. Simulations were used 
to evaluate the method sensitivity, as described in the next section.

\section{Simulations}
We selected for simulation the following parameters:
\textbf{(i)} neutrino spectrum\cite{1}: T= 3.0 MeV, $\eta = 3.0 $; 
\textbf{(ii)} $E_{tot}^{sim} = 3.0 \times 10^{53}$ erg;
\textbf{(iii)} $D_{sim}$ = 10 kpc. 
As a first case study we have taken the LVD experiment \cite{3}, running at the Gran Sasso underground laboratory (Italy) with 
1 kton of active scintillator mass, $~ 1.7 \times 10^{32}$ target protons. The energy threshold was taken as $E_{th} = 4 MeV$. 
For simplicity, we have made $\epsilon(E)=1$.\par

$\bullet$ Simulation steps: \textbf{1)Statistical fluctuations:} $N_\nu$ is calculated from Eq.\ref{numeventos}, and then a random number is draw from a 
Poissonian distribution with mean $N_\nu$, obtaining a new number of events $N_{sf}$ \textbf{; }
\textbf{2)Simulation of experimental $\bar{\nu_e}$ spectrum:} $N_{sf}$ values of energy are then draw following the distribution given by the 
integrand of Eq.\ref{numeventos} \textbf{;} 
\textbf{3) Experimental uncertainties:} Each energy value from step (2) was properly fluctuated accordingly the energy resolution of the 
experiment\cite{4}\textbf{;}
\textbf{4)Systematics induced by uncertainties:} Events bellow energy threshold are discarded\textbf{;}
\textbf{5)End of the data set modelling:} The resulting set of values are used as simulated experimental events to work as the neutrino 
energy spectrum\textbf{.}

\begin{figure}[th]
\centering
    \includegraphics[width=0.88\textwidth]{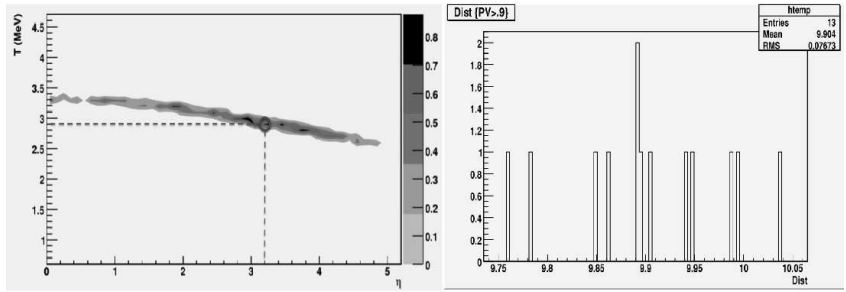}
\caption{Left: KS-probability contours in the parameter space (T,$\eta$). The circle and dashed lines highlights (T,$\eta$)$^*$. 
Right: Distribution of distances D found as solutions of Eq.\ref{numeventos} only
for simulated data with KS-probability greater than 0.9.}
\label{cont-dist}
\end{figure}

$\bullet$ Analysis: In the proposed method, the KS-test is applied in a non-orthodox way: we scan a reasonable range for T and $\eta$ in very small steps, assuming each 
scanned pair (T, $\eta$) as a null hypothesis for the neutrino emission, $(T, \eta)\rightarrow(T,\eta)_0$ . 
The (T, $\eta$) with greater KS probability pointed from the scan is chosen as the “best parametrization” (T,$\eta$)$^*$. 
The corresponding values $N_{sf}$, (T,$\eta$)$^*$, and the assumption for the analysis $E_{tot}^{ana} = 3,0 \times 10^{53}$ erg, are then used 
to solve Eq.\ref{numeventos}, as described before to obtain D. The Fig.\ref{cont-dist} shows in plane (T, $\eta$) 
the results for KS probability, obtained from the simulated spectra. Regardless the uncertainties that were introduced, 
only a small region of the parameter space has significant probability to be recognized as neutrino emission parameters. Moreover, 
the region's boundary suggests a parameter correlation, explained by the mutually inverse 
action of parameters T and $\eta$ in the shape of the neutrino spectrum. In our study the best 
parametrization was found at (T,$\eta$)$^*$ = (2.9, 3.1). The distribution of distances we found solving Eq.\ref{numeventos} by selecting only 
the spectral parameters with KS-probability $> 0.9$ is also shown in figure \ref{cont-dist}. Despite the fluctuations that were introduced we have 
accurately determined the distance $<D> = 9.90 \pm 0.02$ kpc. We also tested the opposite limit, simulating the SN explosion with 
$E_{tot}^{sim}=2.0 \times 10^{53}$ erg but still assuming $E_{tot}^{ana}=3.0  \times 10^{53}$ erg in the simulated data analysis, 
resulting in $<D> \sim 12.6$ kpc.

\section{Conclusions}
The method to estimate the SN distance by neutrino spectral shape is very promising and was tested through simulations. 
When $E_{tot}^{sim}=E_{tot}^{ana}$, we have obtained errors $\Delta D \sim 1\%$, 
meanwhile by testing over extreme limits, i.e. simulating the SN explosion with the lowest plausible energy and still assuming in the simulated 
data analysis the maximum plausible value for $E_{tot}$ (see text for details) we have obtained errors $\Delta D \sim 20\%$.
Enhancements on the simulations should be done by including different parameterization 
of neutrino spectra and more details on experimental uncertainties. Studies on the statistical distribution
of $E_{tot}$ should also be considered to obtain a more robust evaluation of the sensitivity of the method.

\end{document}